\documentclass[12pt]{iopart}

\usepackage{graphicx}% Include figure files
\usepackage{dcolumn}% Align table columns on decimal point
\usepackage{citesort}
\newcommand{\phii}{\phi_{B}^{IV}} % Several macros for symbols
\newcommand{\phic}{\phi_{B}^{CV}}
\newcommand{\rich}{A^{\ast}}

\begin{document}

\title[Co/n-Ge Schottky contacts]{Physical properties of Co/n-Ge Schottky contacts}

\author{L Lajaunie$^{1,2}$, M L David$^1$ and J F Barbot$^1$}

\address{$^1$ Institut Pprime, CNRS-Universit\'e de Poitiers-ENSMA, UPR 3346, bd M.\&P. Curie, BP 30179, 86962 Futuroscope-Chasseneuil cedex, France.}
\address{$^2$ Institut des Mat\'eriaux Jean Rouxel (IMN), Universit\'e de Nantes-CNRS, UMR 6502,  2 rue de la Houssini\`ere, BP 32229,  44322 Nantes Cedex, France.}
\ead{luc.lajaunie@cnrs-imn.fr}

\begin{abstract}
To investigate the role of the interface state on the physical properties of Schottky contacts, Co/n-Ge Schottky diodes that have undergone various cleaning methods (HF etching and \textit{in-situ} thermal cleaning) were studied by Transmission Electron Microscopy (TEM), Deep-level Transient Spectroscopy (DLTS) and by a detailed analysis of the temperature dependence of the diodes characteristics. It is shown that Schottky barrier height characteristics are sensitive to the nature of the interface. The strongest Fermi level pinning and the highest spatial inhomogeneities are observed for intimate metal/semiconductor contacts. The presence of a thin oxide interlayer, even of Ge native oxide, allows the Fermi level to be released towards the conduction band and leads to more homogeneous contacts. Finally our results suggest that a pure $GeO_2$ oxide interlayer should present a better depinning efficiency than the native Ge oxide.
\end{abstract}

%Uncomment for PACS numbers title message
% Keywords required only for MST, PB, PMB, PM, JOA, JOB? 
%\vspace{2pc}
%\noindent{\it Keywords}: Article preparation, IOP journals
% Uncomment for Submitted to journal title message
\submitto{\JPD}
% Comment out if separate title page not required
\maketitle

\section{\label{intro}Introduction}

Owing to its high charge carrier mobility and its good compatibility with high-k materials,\cite{dimoulas07a,kamata08MT} germanium is a potential candidate to replace silicon as channel material in sub-22 nm Complementary Metal Oxide Semiconductor (CMOS) technologies. However  Ge n-MOSFETs (Metal Oxide Semiconductor Field Effect Transistors) present lesser performance than silicon n-MOSFET mostly because of the fast diffusion of n-type dopants in germanium which prevents the formation of shallow junctions.\cite{brotzmann08} To overcome such issue Schottky Source/Drain MOSFETs are promising alternatives.\cite{kobayashi09} However whereas such devices require low Schottky barriers (SBHs) to compete with conventional MOSFETs,\cite{dubois04} most of the direct contacts metal/Ge yield to high SBHs due to a strong Fermi level-pinning effect.\cite{dimoulas09mrs} SBHs larger than the germanium band gap have even been reported suggesting the formation of an inversion layer at the interface metal/Ge\cite{chi05} consecutively of the small germanium band gap and of the charge neutrality level lying close to the valence band.\cite{tsipas09} Promising studies have shown that by introducing a thin oxide layer at the metal/Ge interface, the Fermi level (FL) can be depinned,\cite{zhou08,kobayashi09} i.e the FL of the metal is released toward the conduction band of germanium, thus yielding a lower SBH. For instance, by inserting a thin layer (2nm) of $Al_2O_3$ at the metal/semiconductor (MS) interface, SBHs of Co/n-Ge diodes has been found to decrease from 0.62 to 0.35 eV.\cite{zhou08} Moreover it has been shown that the depinning efficiency of the oxide is not only dependent on the nature of the metal,\cite{zhou08} but also on the thickness and on the chemical nature of the oxide\cite{kobayashi09}.
The underlying mechanisms leading to the FL pinning are still subject to debate. According to the metal induced gap states (MIGS) theory, the FL pinning is due to the wave function of the metal penetrating into the semiconductor and inducing additional states in the band gap. \cite{dimoulas09mrs} However the role of the dangling bonds at the MS interface has been pointed out as well.\cite{tung00}
In the case of germanium, the different experimental observations tend to strengthened one or the other theory.\cite{zhou10,kobayashi09,nishimura07,yamane10} However, in most of the case, possible SBHs non-homogeneities which can alter the electrical properties of real MS contacts\cite{werner91,zhu00,pakma2008,kavasoglu09} are not taken into account.

Surface preparation of the substrate prior to metal deposition could be a possible source of SBH inhomogeneities. In the studies concerning the characterization of SBHs on germanium, HF etching is the most often used cleaning method.\cite{zhou08,husain09a} As the roughness of the germanium surface increases by a factor two with this cleaning method \cite{sun06} it could enhance the spatial inhomogeneities at MS interfaces. Moreover the metal impurities at the germanium surface are not removed with such a cleaning in contrary to what is observed after an HCl cleaning for instance. \cite{kamata08} This may also influence the SBH characterization. Another way to clean the germanium surface is to realize the thermal decomposition of the native oxide layer $GeO_X$ as its desorption occurs at 430$^\circ$C under high vacuum.\cite{prabhakaran00} An annealing in UHV at 360$^\circ$C during 15 minutes has even been reported to be sufficient to obtain an oxide-free germanium surface\cite{dimoulas06}. Besides the modifications that can be induced by different surface cleaning prior to MS contact formation, implantation induced defects created during the metal deposition also modify the properties of MS contacts.\cite{lauwaert09,auret06}

In this work, electrical and microstructural characterizations of Co/n-Ge contacts have been conducted to understand the influence of the interface state on the electrical properties of MS contacts. In particular a detailed analysis of the temperature dependence of the diodes characteristics has been carried out to take into account the inhomogeneities of the SBHs. This paper is divided into six parts. In the first part,  the theoretical background related to SBH inhomogeneities analysis is recalled. In the second part, the experimental details are given. In parts three to five, the microstructural and electrical characterization of the samples are presented. Finally the results are discussed in the framework of the Fermi level-pinning concept.

\section{\label{bckgd}Background}

Under the assumption of thermionic current being the major forward current distribution, the diode current of a homogeneous MS interface can be determined as: \cite{rhoderick82}

\begin{eqnarray}
I=I_{S}(e^{\frac{q(V-IR_S)}{nkT}}-1)
\label{eq1}
\end{eqnarray}

where V is the applied voltage, q the elementary charge, n the ideality factor, k the Boltzmann's constant, T the temperature and $R_S$ the series resistance. The saturation current, $I_S$, is given by: 

\begin{eqnarray}
I_{S}=AA^{\ast}T^{2}e^{\frac{ -q\phii }{kT}}
\label{eq2}
\end{eqnarray}

where A is the Schottky contact area, $\rich$ the Richardson's constant and $\phii$ the SBH derived from the I-V characteristics. The image force lowering has been neglected due to the low doping of the germanium. The SBH can also be estimated from C-V measurements using the following relation: 

\begin{eqnarray}
q\phic=q(V_{bi}+\frac{kT}{q}+\frac{kT}{q}\ln{\frac{N_C}{N_D}})
\label{eq4}
\end{eqnarray}

where $N_C$ is the effective density of states in the conduction band, $V_{bi}$ the built-in potential and $N_D$ the free carrier concentration.

Most of the interpretations of electrical data from real Schottky contacts implicitly assume the uniformity of the SBH at the MS interface. However, several discrepancies with the thermionic emission theory such as differences between $\phii$ and $\phic$, variation of the ideality factor with the temperature as well as the non-linearity of the Richardson's plot are often noted in the literature.\cite{zhu00,pakma2008,kim09,mtangi09,biber03} Fluctuations of the built-in potential ascribed to spatial inhomogeneities along the MS interface alter the electrical behavior of real Schottky contacts. These fluctuations are not taken into account in the thermionic (TE) model but are well reproduced by the model of Werner and G{\"u}ttler \cite{werner91} (the WG's model in the following) by assuming a Gaussian distribution of the SBH of mean value $ \overline{\phi_B}$ and standard deviation $\sigma$. The authors have shown that the SBH determined from C-V measurements is not dependent on the standard deviation, i.e. that $\phi_{B}^{CV}(T)=\overline{\phi_B(T)}$. Eqs (\ref{eq1}) and (\ref{eq2}) giving the flow of current in the TE model are similar than in the WG's model, but with an apparent SBH given by:

\begin{eqnarray}
\phi_{B}^{IV}(T)&=\overline{\phi_B(T)}-\frac{q\sigma^2(T)}{2kT}  
\label{eq7}
\end{eqnarray}

Both, $\overline{\phi_B}$ and $\sigma^2$ are temperature dependent according to:\cite{werner91}

\begin{eqnarray}
\overline{\phi_B(T)}&=\overline{\phi_0}+\alpha_\phi T \label{eq8} \\
\sigma^2(T)&=\sigma_0^2+\alpha_\sigma T
 \label{eq9}
\end{eqnarray}

where $\overline{\phi_0}$ and $\sigma_0$ are the mean barrier height and the standard deviation extrapolated at 0 K; $\alpha_\phi$ and $\alpha_\sigma$ are their temperature coefficients, respectively. $\overline{\phi_0}$ and $\alpha_\phi$ are determined by plotting $\phic$ as a function of the temperature. The standard deviation at 0 K and its temperature coefficient are obtained by combining equations (\ref{eq7}) and (\ref{eq9}):

\begin{eqnarray}
\phic(T)-\phii(T)&=\frac{q\sigma_0^2}{2kT}+\frac{q\alpha_\sigma}{2k}
\label{eq10}
\end{eqnarray}

The conventional Richardson's plot is then modified as follows to take into account the SBH inhomogeneities:

\begin{eqnarray}
\ln{\frac{I_{S}}{T^2}}-\frac{q^2\sigma^2(T)}{2k^2T^2}&=\ln{(A \rich e^{\frac{-q\alpha_\phi}{k}})}-\frac{\overline{q\phi_0}}{kT}
\label{eq11}
\end{eqnarray}

The Richardson's constant, $\rich$, and the mean SBHs at 0 K, $\overline{\phi_0}$, are extracted from the modified Richardson plot, Equation (\ref{eq11}). In the following we have labeled this last value $\overline{\phi_0^{RICH}} $ for the distinction with the value deduced from the C-V-T measurements ($\overline{\Phi_0}(T)$, Equation (\ref{eq8})).

In the WG's model, the temperature dependence of the ideality factor is of importance as it could modify the extracted parameters of the Gaussian distribution. This dependence is related to the deformation of the Gaussian distribution under the applied bias: \cite{werner91}

\begin{eqnarray}
\frac{1}{n}-1&=- \rho_2+\frac{\rho_3}{2kT}
\label{eq12}
\end{eqnarray}

where $\rho_2$ and $\rho_3$ quantify the voltage deformation of the mean barrier height and the standard deviation, respectively. As n reflects the deformation of the Gaussian distribution under the applied bias, the SBH extracted from I-V characteristics is already corrected from the voltage deformations. However, any voltage dependencies of the mean barrier height will impair the comparison between $\phic(T)$ and  $\phii(T)$ which is needed to deduce the standard deviation of the Schottky contacts. From this point it is assumed in the WG's model that the mean SBH shows under reverse bias the same dependence than under forward bias. Then, the correct zero-bias barrier, $\overline{\phi_{B0V}}$, extracted from C-V characteristics is given by:

\begin{eqnarray}
\overline{\phi_{B0V}}=V_{bi}(1-\rho_2)+\frac{kT}{q}+\frac{(E_C-E_F)}{q}
\label{eq15}
\end{eqnarray}

The free carrier concentration has then to be corrected accordingly: 

\begin{eqnarray}
N_{D0V}=N_D(1-\rho_2)
\label{eq16}
\end{eqnarray}

It must be pointed out that this later assumption has never been experimentally confirmed even if some of these corrections have already been used in the literature. \cite{wenckstern06}

\section{\label{exp}Experiments}

N-type (001) germanium wafers ($\rho \sim$ 20  $\Omega$.cm) were used in this study. After different surface cleaning (pre-treatment) that will be detailed in the following, cobalt films of 22 nm thick were deposited at room temperature by using an electron beam deposition system under high vacuum ($10^{-8}$ torr). Thermal pre-treatments were performed \textit{in-situ} in the deposition chamber using four lamps of 2000 W located in the vicinity of the sample holder. For each run, two samples were mounted in the deposition chamber. One of the samples was dedicated to the structural characterization and the film was deposited on all the sample surface. For the other one, the Co film was evaporated through a circular shadow mask of 2 mm in diameter to allow the electrical study of the Co-based Schottky contacts. No possible effects related to the perimeter of the diodes have been highlighted by a variable-area diode preliminary study.

Two different surface pre-treatments were studied: HF etching and thermal cleaning. For HF etching, the sample labeled PHF was dipped successively in acetone, ethanol solutions, HF (4\%) during 30 seconds, and then rinsed three times in deionised water. Finally it was dried under $N_2$ air before being loaded in the deposition chamber. The characteristic time of passivation of a Ge surface being of the order of 15 min,\cite{pelissier07} a great care was taken to load the samples in the deposition chamber as fast as possible  knowing that the usual time to load the samples in the deposition chamber is about 5 min. For the thermal cleaning: two temperatures were chosen: 400$^\circ$C and 700$^\circ$C (samples labelled P400 and P700). The duration and the ramp were set at 40 min and 20 min, respectively. Note that no chemical treatment was performed on the surface of the samples before any thermal pre-treatment. Finally a reference sample, labelled REF, which was not submitted to any pre-treatment, was also studied.

TEM experiments were carried out to characterize the nature of the metal/semiconductor interface and the microstructure of the Co thin films. TEM samples were prepared in the cross-section geometry; they were mechanically thinned using a tripod polisher down to 10 $\mu$m, and then ion milled in a GATAN-PIPS apparatus at low energy (2.5 keV Ar) and low incidence ($\pm$8$^\circ$) to minimize irradiation damage. They were studied by High Resolution Transmission Electron Microscopy (HRTEM) using a JEOL 3010 microscope (300 kV, $LaB_6$, point to point resolution=0.19 nm). The electrical characterizations (C-V, I-V, DLTS) of the diodes were performed using a BioRad DL8000 (bridge capacitance at 1 MHz) apparatus with a He cryostat allowing a temperature variation from 40 to 350 K with a temperature sensitivity of 0.1 K. Moreover, optical microscopy images were used to accurately determine the area of each diode.

\section{\label{sec:level1}Microstructural characterization}

\begin{figure}[!ht]
\includegraphics[width=9cm]{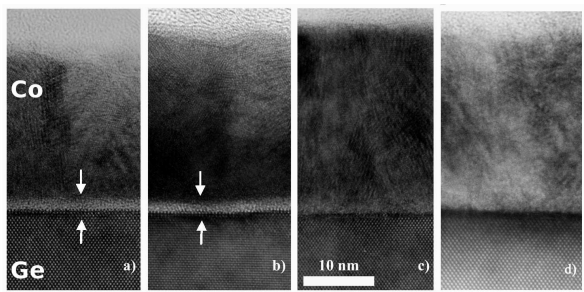}
\caption{TEM bright-field micrographs of Co thin films on Ge substrate deposited by electron beam evaporation. a) $REF$ b) $P400$ c) $P700$ d) $PHF$. The white arrows highlight the oxide interlayer.}
\label{fig1}
\end{figure}

Figure \ref{fig1} displays cross-sectional HRTEM micrographs of the four studied samples after cobalt deposition. At the top of the samples, a continuous layer of polycrystalline cobalt, 22-25 nm thick, with nanometer sized grains is observed. According to selected area electron diffraction patterns analyses (not shown here), for any pre-treatments the cobalt thin films exhibits the hexagonal closed-pack structure as expected\cite{jerl}.  The cleaning procedure has thus no effect on the microstructure of the cobalt film.

Between the cobalt layer and the germanium substrate, a thin amorphous-like layer (2 nm thick) attributed to germanium oxide is observed in both the reference sample (REF) and the sample pre-treated at 400$^\circ$C (P400). No evidence of this amorphous layer is seen on samples pre-treated at 700$^\circ$C (P700) nor by HF (PHF). These observations show that a threshold temperature $T_S$, $400^\circ C<T_S<700^\circ C$, exists, above which the native oxide can be removed under annealing in high vacuum. This temperature is, however, strongly dependent on the atmosphere; the reduction of $GeO_X$ under wet $N_2$ ambient\cite{zou07} was observed at 550$^\circ$C  whereas the desorption in ultra high vacuum of the native germanium oxide was observed by heating the substrate at 360$^\circ$C during 15 minutes\cite{dimoulas06}. The temperature is thus not the only parameter which takes part in the removal of the oxide, and the other parameters, as the pressure inside the deposition chamber, must be taken into account. Finally, although not revealed by HRTEM, the chemical composition of the oxide interlayer could have been modified by the annealing at 400$^\circ$C \cite{prabhakaran00}, leading to different SBHs. This will be confirmed by electrical characterization.

From these HRTEM observations, it is clear that the MS interface is sensitive to the cleaning history of the samples; therefore differences in the Schottky barriers properties are thus expected.  \\

\section{\label{dlts}Deep levels characterization}

\begin{figure}[!ht]
\includegraphics[width=6cm]{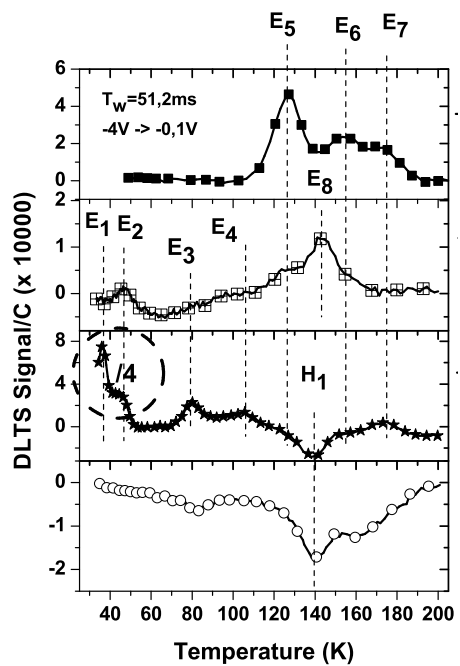}
\caption{DLTS scan normalized by the capacitance for the four as-deposited samples. The amplitude of the encircled area have been divided by four for convenience.}
\label{fig2}
\end{figure}

DLTS measurements were performed on the four samples in order to reveal a possible contamination which would have been able to occur during the preparation of the surface and/or during the deposition.  In Figure \ref{fig2}, the DLTS spectra of the four samples are shown. Several peaks are observed, the energy of the corresponding deep levels with respect to the bottom of the conduction band edge and their apparent capture cross section are reported in Table \ref{tab1}. A hole trap, $H_1$, is observed for the P700 and PHF samples. The observation of minority carrier traps in Ge Schottky barriers has already been reported for Co and Cr n-Ge Schottky barriers.\cite{simoen08d,simoen08} This is attributed to the creation an inversion layer near the surface.\cite{markevich04,simoen08} As it will be detailed in the following, the presence of an inversion layer is also highlighted by the SBHs characteristics of these two diodes.

Most of the signatures could be assigned to metallic impurities. For instance, from their signatures and by comparing with DLTS-spectra measured on samples where copper was intentionally introduced, $E_7$ and $H_1$ could be assigned to the two acceptor levels of copper $Cu_S^{2-/3-}$ and $Cu_S^{-/2-}$, respectively. Moreover, $E_2$ and $E_3$ could respectively be assigned to the two acceptor levels of gold \cite{pearton82} and $E_6$ and $E_4$ to nickel \cite{simoen06} and titanium in substitutionnal position.\cite{lauwaert09} The nature of the other traps $E_1$, $E_5$, $E_8$ is still unclear. Irradiation-induced defects, created by some energetic particle originating from the region of the filament as observed by Auret \textit{et al.} \cite{auret06,auret08} can however be ruled out since the signature of these deep levels does not correspond to any of the electron-irradiation induced traps reported in the literature.\cite{fage00}

\begin{table}[!ht]
\caption{\label{tab1}Signatures of the different DLTS lines on the as-deposited Co/n-Ge diodes. The capture cross section of the hole trap $H_1$ should be considered with caution since it has been calculated from majority carrier (electron) characteristics.* All the energies are given with respect to the conduction band edge except for the hole trap $H_1$, where it is given with respect to the valence band edge. The concentrations are relative to the samples which are not in brackets.}
\begin{tabular}{@{}llllll}
\br
 Sample & Deep Levels						    & $E_{C}-E_{T}$(eV)       & $\sigma(cm^{2})$  & $N_{T}(cm^{-3})$ & Possible identity	\\ \mr
 P700, (P400)& $E_{1}$     & -     & -       & -     &       ? \\
 P700,(P400) &$E_{2}$ & 0.07        		  & $3\times10^{-15}$               & $1.6\times10^{12}$     & Au   \\  
 P700, (P400) & $E_{3}$ & 0.18							& $2\times10^{-13}$   			      & $3\times10^{11}$			&			?		\\
 P700, (P400) & $E_{4}$ & 0.22							& $2\times10^{-15}$   			      & $1.8\times10^{11}$			& $Ti^{-/2-}$						? \\
 REF & $E_{5}$ & 0.21							& $2\times10^{-16}$   			      & $5\times10^{11}$				&		?		 \\
 REF, (P700) & $E_{6}$ & 0.31							& $8\times10^{-15}$   			      & $1\times10^{11}$					&	$Ni^{-/2-}$				  \\
 REF, (P700) & $E_{7}$ & 0.34							& $2\times10^{-15}$   			      & $1\times10^{11}$				& $Cu^{2-/3-}$				  \\
 P400 & $E_{8}$ & 0.31							& $9\times10^{-14}$   			      & $1\times10^{11}$				& ?				  \\
 PHF, (P700) & $H_{1}$ & 0.31*							& $6\times10^{-14}$   			      & -				&			$Cu^{-/2-}$	  \\
 \br
\end{tabular}
\end{table}

The total defect concentration is below $5\times10^{12}$ $cm^{-3}$, so 5\% of the doping concentration in the region probed (3.5-10 $\mu m$ from the surface). Due to the low defect concentration, it has been difficult to obtain reliable defect profiles. However, one can estimate from the high diffusion coefficients of the metals detected in these samples \cite{claeys07} that the defect concentration closer to the surface is not going to reach a level where it can affect the I-V forward characteristics neither by generation-recombination nor by compensation. For Nickel which is known to be the most efficient lifetime killer in germanium, a concentration larger by several orders of magnitude would be necessary to influence the I-V forward characteristics.\cite{simoen07} The concentration of defects is not significant. Indeed, it varies for diodes having undergone the same procedure,  suggesting a random contamination during either the pre-treatement or the deposition. Copper is a common contaminant in germanium \cite{simoen08d,simoen07,simoen08} but it is still unclear whether the Cu-contamination is coming from the wafer itself or from the co-sputtering of copper during the metal deposition. \cite{simoen08} On the other hand, $E_1$, $E_2$, $E_3$ and $E_4$ are only observed in the pre-annealed samples P400 and P700, with a higher concentration in P700 than in P400. This clearly shows that a high temperature pre-treatment introduces metallic impurities, the higher the temperature, the higher the concentration of impurities. This contamination could be due to impurities located in the deposition chamber, that desorb and then diffuse in the germanium substrate during the pre-annealing.

In the following, due to the low concentration of deep-levels, the generation-recombination current has been neglected and the current flow has been considered as predominantly thermionic. 
 
 \section{\label{sbh}Schottky barrier characterization }

\subsection{\label{160k}Electrical characterization at 160 K}

\begin{figure}[!ht]
\includegraphics[width=8cm]{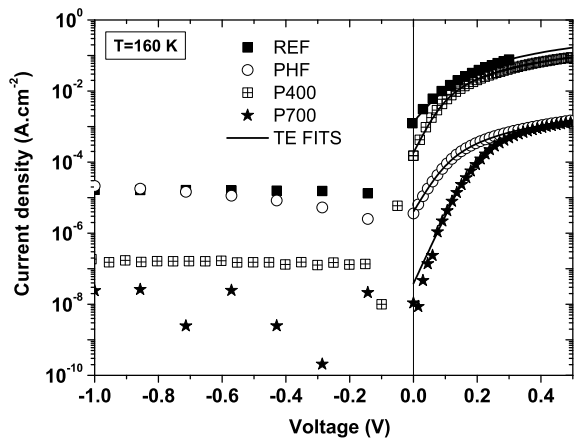}
\caption{Reverse and forward I-V currents at 160 K for the four Co/Ge contacts. Fitting curves using the thermionic model (TE) are in plain lines.}
\label{fig3}
\end{figure}

To minimize the thermal generation of intrinsic carriers ($n_i$$\sim$ $10^{13}$ $at/cm^3$ at 300 K), the four diodes were studied at a given temperature of 160 K. Figure \ref{fig3} shows typical semi-logarithmic I-V plots of the four diodes measured at this temperature.

As seen, for each pre-treatment, a good quality rectifying Schottky barrier is observed with several orders of magnitude between the forward and the reverse currents. However, depending on the pre-treatment, different behaviours are observed. For instance, the REF and P400 diodes exhibit much higher forward currents than the two other diodes. The P700 diode presents the lowest reverse current, lower than the limit current of our measurement set-up. The forward currents were fitted with the TE model (plain lines on Figure \ref{fig3}) which was assumed to be the dominant transport mechanism for such low doping concentration. As seen, a good agreement between experimental and theoretical data is observed. The parameters, namely the saturation current $I_S$, the Schottky barrier height $\phii$, the ideality factor n and the series resistance $R_S$ were extracted from the TE fits for each diode.  The as-derived series resistance and ideality factors were also confirmed by using the plots of experimental current over conductance (I/G) as a function of I (not shown here). \cite{werner88} From the C-V measurements at 160 K, the doping concentration was found to be $(8.6 \pm 0.5)  \times 10^{13}$ $cm^{-3}$ in good agreement with the manufacturer specification ($9.0\times10^{13}$ $cm^{-3}$). The main parameters extracted at 160K are summarized in Table \ref{taba}. \\

\begin{table}[!ht]
\caption{\label{taba} SBHs extracted from C-V measurements $\phic$, from I-V measurements $\phii$ and ideality factors $n$ at 160 K.}
\begin{indented}
\item[]\begin{tabular}{@{}llll}
\br
Sample	  & $q\phic$ ($eV$)  & $q\phii$ ($eV$)  & $n$ \\
\mr
 REF &  0.34 & 0.30 & 2.8 \\
 PHF &  0.85 & 0.38& 2.0 \\
 P400 &  0.63 &0.33  & 1.8 \\
 P700 &  0.92 & 0.44 & 1.7 \\
 \br
\end{tabular}
\end{indented}
\end{table}

All the diodes present ideality factors greater than unity. The SBHs both extracted from C-V and I-V measurements are found to be strongly dependent on the pre-treatment procedure. Furthermore, for all the diodes, the SBH derived from C-V measurements is much higher than that derived from I-V measurements ($q\phic>q\phii$).

\subsection{\label{80270}Electrical characterization in the temperature range 80 - 270 K}

\begin{figure}[!ht]
\includegraphics[width=9cm]{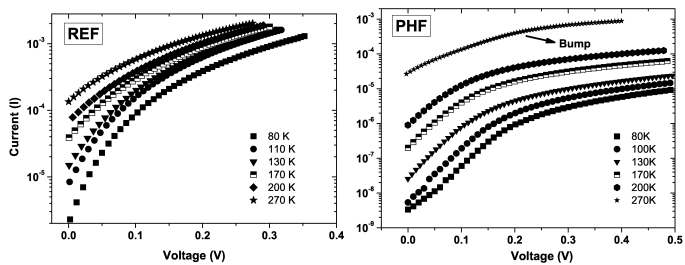}
\caption{Forward I-V characteristics at different temperatures for the two diodes REF and PHF. The arrow highlights the presence of a second transport mechanism above 220 K.}
\label{fig4}
\end{figure}

To get further insights on the carrier transport mechanisms through the Co/n-Ge contacts, I-V and C-V characteristics were recorded in the temperature range 80 - 270 K every 10 K for each samples. Figure \ref{fig4} shows the forward I-V characteristics of the two diodes REF and PHF which present similar behaviors with the P400 and P700 diodes, respectively. An increase of the current with the temperature is observed in good agreement with the thermionic model. It is noteworthy that from a temperature of 220 K, a small bump on the PHF and P700 diodes I-V characteristics is observed, pointing out clearly the presence of a second transport mechanism.  To get ride of this mechanism, the study will be limited to the low SBH (low temperature side). \\

\begin{figure}[!ht]
\includegraphics[width=6cm]{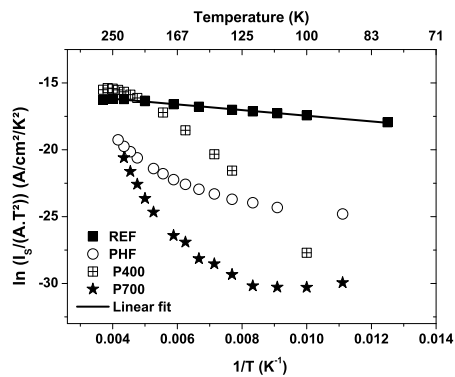}
\caption{Conventional Richardson's plots of Co/Ge contacts having undergone various pre-treatments.}
\label{fig5}
\end{figure}

The saturation currents were extracted from the I-V curves and used to plot the conventional Richardson plots shown in Figure \ref{fig5}. As seen, the expected linear behavior is only observed for the REF diode. However, the SBH and the Richardson constant derived from this plot are abnormally low: 0.02 eV and $2\times 10^{-7}$ $A.cm^{-2}K^{-2}$ respectively. This later value is several order of magnitude lower than the theoretical value of $\rich$ for n-(001) Ge (143 $A.cm^{-2}K^{-2}$ in reference \cite{sze07}).

\begin{figure}[!ht]
\includegraphics[width=6cm]{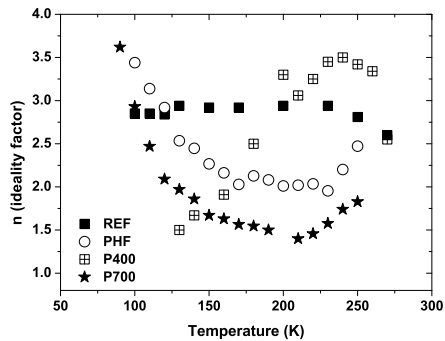}
\caption{Ideality factor, n, extracted from forward I-V characteristics as a function of the temperature for the four Co/Ge contacts.}
\label{fig6}
\end{figure}

The values of n determined from the fit procedure are plotted in Figure \ref{fig6}. As seen, the ideality factor is strongly dependent on temperature on the low temperature side (below 200 K) except for the REF diode where it stays constant on all the temperature range studied.

These observations conjugated with the differences between $\phic$ and $\phii$ at 160 K are good indicators of SBH inhomogeneities. To successfully apply the WG's model, a temperature range of application has been determined by considering that \textit{i.)} the current flow must be governed by a single thermionic mechanism and \textit{ii.)}  $\phic$ must be larger than or equal to $\phii$ (see Equation (\ref{eq10})).  From these two points, the temperature range has been restricted below 180 K.

\subsection{\label{gauss}Analysis of barrier inhomogeneities using a Gaussian distribution model}

\begin{figure}[!ht]
\includegraphics[width=6cm]{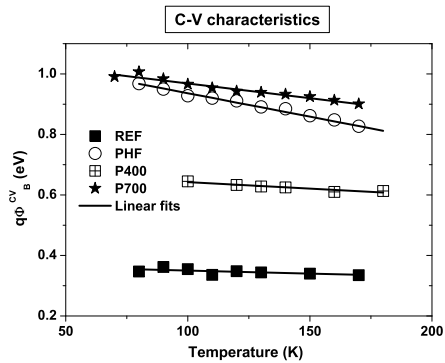}
\caption{Variation of $\phic$ with the temperature and linear fits from Equation (\ref{eq8}).}
\label{fig7}
\end{figure}

\begin{figure}[!ht]
\includegraphics[width=6cm]{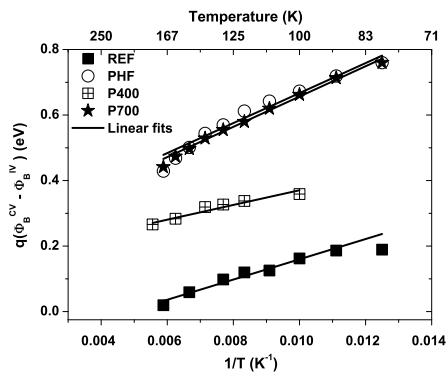}
\caption{Variation of $\phic(T)-\phii(T)$ with $\frac{1}{T}$ and fitting curves from Equation (\ref{eq10}) for the as-deposited samples.}
\label{fig8}
\end{figure}

Figure \ref{fig7} and Figure \ref{fig8} show the SBH extracted from C-V-T measurements and $\phic(T)-\phii(T)$ as a function of temperature, respectively. As seen, these plots are straight lines and $q\phic$ higher than $q\phii$ is observed in all the temperature range. The derived parameters from the linear fits, namely the mean barrier height at 0 K $\overline{\phi_0}$, its temperature coefficient $\alpha_\phi$, the standard deviation at 0K $\sigma_0$, and its temperature coefficient, $q\alpha_\sigma$,were extracted and are reported in Table \ref{tab3}.

\begin{figure}[!ht]
\includegraphics[width=6cm]{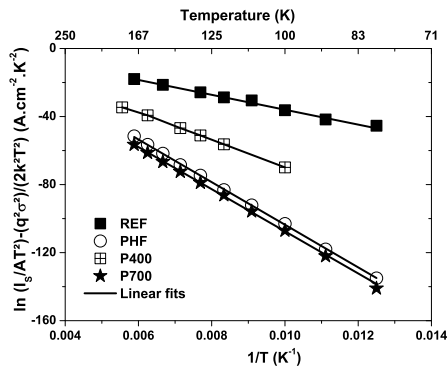}
\caption{Modified Richardson's plot and linear fits from Equation (\ref{eq11}). To minimize the uncertainty on the determination of both $\sigma_0$ and $\alpha_\sigma$, the term $\frac{\sigma^2(T)}{2k^2T^2}$ has been determined by using the experimental $\phic(T)-\phii(T)$ data.}
\label{fig9}
\end{figure}

The modified Richardson plot according to Equation (\ref{eq11}) is plotted in Figure \ref{fig9}. As seen, the expected linear behavior is recovered after the WG's model corrections. The Richardson's constant and the mean barrier height at 0 K ($\overline{\phi_0^{RICH}}$) were derived as explained in previous section and are reported in Table \ref{tab3}. The Co/n-Ge diodes prepared by using different cleaning procedures show significantly different SBH properties. Furthermore, for all the diodes, the values of $\rich$ are close to the theoretical value. In addition, the mean SBHs at 0 K determined using the modified Richardson plots, $\overline{\phi_0^{RICH}}$, are in good agreement with those extracted from the C-V-T characteristics, $\overline{\phi_0}$. These two observations confirm that the experimental data are well described by a Gaussian distribution of SBH.

\begin{figure}[!ht]
\includegraphics[width=6cm]{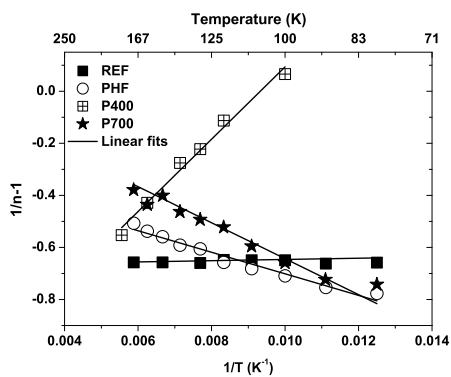}
\caption{$n^{-1}-1$ vs $T^{-1}$ for the as-deposited samples and the linear fits from Equation (\ref{eq12}).}
\label{fig10}
\end{figure}

In this temperature range (80-170 K), in good agreement with the WG's model, a plot of 1/(n-1) versus 1/T yields to a straight line as shown in Figure \ref{fig10} for all the diodes. \\

From these plots, the voltage coefficients $\rho_2$ and $\rho_3$ were extracted and reported in Table \ref{tabv}. As seen, they are strongly dependent on the cleaning history. Table \ref{tabv} lists also the experimental free carrier concentration as well as the corrected free carrier concentration $N_{D0V}$ at 160 K. Not only this correction cannot be applied to the P400 diode (as $\rho_2>1$) but it leads also to an increase of the discrepancy of the free carrier concentration without any physical reason. Obviously, the mean SBH of samples $\overline{\phi_B}$ does not show under reverse bias the same dependence than under forward bias;  therefore no further correction are needed on the SBH extracted from C-V characteristics.

\begin{table}[!ht]
\caption{\label{tabv}Voltage coefficients, $\rho_2$ and  $\rho_3$ deduced from Figure \ref{fig10}. Free carrier concentrations extracted from C-V characteristics at 160 K without ($N_D$) and after corrections of the voltage deformation ($N_{D0V}$).}
\begin{indented}
\item[]\begin{tabular}{@{}lllll}
\br
Sample	  & $\rho_2$  & $\rho_3$ ($meV$) & $N_D$ $(cm^{-3})$ & $N_{D0V}$  $(cm^{-3})$ \\
\mr
 REF &  0.67 & 0 & $8.3 \times 10^{13}$ & $2.7 \times 10^{13}$ \\
 PHF &  0.29 & -7 & $9.1 \times 10^{13}$ & $6.5 \times 10^{13}$ \\
 P400 &  1.30 & 24  & $8.1 \times 10^{13}$ & X   \\
 P700 &  -0.05 & -1 & $9.6 \times 10^{13}$ & $1.1 \times 10^{14}$ \\
 \br
\end{tabular}
\end{indented}
\end{table}

\section{\label{disc}Discussion: SBHs properties}

\begin{table*}[!ht]
\caption{\label{tab3}Mean barrier height at 0 K $ \overline{\phi_0} $, and its temperature coefficient $\alpha_\phi$,  Standard deviation at 0 K $\sigma_0$ and its temperature coefficient $\alpha_\sigma$. Richardson's constant, $\rich$, the mean barrier height at 0 K derived from the modified Richardson plot, $\overline{\phi_0^{RICH}} $. }
\begin{indented}
\item[]\begin{tabular}{@{}lllllll}
\br
&$q \overline{\phi_0} $& $q\alpha_\phi$& $q\sigma_0$& $q\alpha_\sigma$& $\rich$ & q$\overline{\phi_0^{RICH}} $\\
sample& (eV)&(meV/K)&(meV)& (meV$^2$/K) &(A.cm$^{-2}$.K$^{-2}$)&(eV)\\
\mr
 REF			& 0.37   &  -0.21   	& 73   &  -26  	&  143  &  0.38      \\
 PHF      & 1.09   &  -1.55   & 89   &  36  &  101  & 1.09  \\
 P400     & 0.69   &   -0.42  & 62   &  25   &  59  &  0.68    \\
 P700     & 1.07   &   -0.98  & 89   &   34   &  136 &  1.07  \\
 \br
\end{tabular}
\end{indented}
\end{table*}

According to Table \ref{tab3}, all the surface pre-treatments lead to both an increase of the mean SBH at 0 K, $\overline{\phi_0}$, and of its temperature coefficient, $\alpha_\phi$. The PHF and P700 diodes, both being intimate contacts, present similar values of $\overline{\phi_0}$. The passivation of the dangling bonds by HF pretreatment has thus no or few effects on the FL pinning. The presence of a thin oxide interlayer (REF and P400 samples) leads to the decrease of $\overline{\Phi_0}$. However, the P400 sample exhibits a 80 \% higher mean SBH at 0 K than the REF sample for equal oxide thickness. This clearly points out the different chemical nature of the oxide between the P400 and the REF sample. According to the experiments of Prabhakaran \textit{et al.},\cite{prabhakaran00} the oxide interlayer of the REF sample is made of $GeO_2$ and other suboxides while it mainly consists of suboxides for the P400 sample. The oxide thicknesses of both REF and P400 are similar, this shows that the depinning efficiency is higher for $GeO_2$ than for other suboxides. Finally it is worth noticing that the PHF and P700 samples exhibit a higher mean SBH than the germanium band gap in the temperature range studied (see Figure \ref{fig7}), as already reported in previous experimental and theoretical studies on n-type germanium. This is attributed to the FL pinning and to the formation of an inversion layer \cite{chi05,yao06,tsipas09} thus explaining the observation of a minority carrier trap for these two diodes.

The temperature dependence of the mean SBH ($\alpha_\phi$) can also be roughly explained in the framework of the Fermi level pinning concept.\cite{werner93,zhu00} According to the band to which the FL is pinned, the temperature coefficient varies from 0 to $dE_g/dT$, where $E_g$ is the indirect band gap of germanium. In the temperature range 100-300 K, the value of $dE_g/dT$ for germanium is close to (-0.4)-(-0.5) meV/K.\cite{auvergne74,lautenschlager85} The temperature coefficient of the REF diode thus indicates a pinning between the conduction band and the middle of the band gap. However, for high SBHs, $q\alpha_\phi$ is four times higher than the expected value and therefore the FL pinning is not sufficient to understand the variation of SBH with the temperature.

The standard deviation at 0 K is a measure of the barrier inhomogeneity: the higher the value $\sigma_0$, the higher the SBH inhomogeneity. The highest values of $\sigma_0$ are found for the PHF and P700 diodes, indicating thus, that the intimate contact of Co on Ge enhances the inhomogeneity of the SBH distribution. This result is of importance as it points out the need to take into account the SBH inhomogeneities even if no interfacial layer is observed at the MS interface.  Inhomogeneous SBH can be induced by the roughness of the MS interface \cite{werner91}. It is well known that a HF pre-treatment yields an increase of the roughness of the Ge substrate \cite{sun06,rivillon05} thus explaining the higher $\sigma_0$ observed for the $PHF$ diode than for the REF and P400 diodes. Even if no study on the impact of the Ge roughness by thermal pre-treatments has been found in the literature, similar statement can be derived for the $P700$ diodes according to the value of $\sigma_0$. 
Finally, it should be noted that, contrary to what was observed for $\overline{\phi_0}$, the standard deviation is less affected by the chemical nature of the oxide interlayer. 

The temperature coefficient, $\alpha_\sigma$, is strongly modified by the nature of the pre-treatment as well. The temperature coefficients are found to be negative for the REF diode, and superior to 0 for the PHF, P400 and the P700 diodes, indicating thus that the inhomogeneity increases with the temperature. According to Zhu \textit{et al.},\cite{zhu00} the physical meaning of the negative value of $\alpha_\sigma$ can be understood by the pinch-off model proposed by Tung.\cite{tung92} In this model, the variation of SBH between low and high SBH areas decreases with the temperature. The barrier heights appear more homogeneous than they are. Therefore, the dependence with the temperature of the standard deviation could not be understood by the pinch-off model for the P400 and the P700 diodes. Such behavior, according to us, has never been reported in the literature.

\section{\label{conc}Conclusion}

In summary, by conducting together microstructural and electrical characterizations of Co/n-Ge diodes that have undergone various cleaning methods, we have linked the SBH properties to the interface state. For any pre-treatments, the electrical behaviors of the Schottky contacts could not be explained by the thermionic model only, pointing out the necessity to think in term of barrier inhomogeneity. The electrical measurements can be satisfactorily explained by assuming a Gaussian distribution of SBH as already reported for Ni/n-Ge Schottky contacts.\cite{li09} Whatever the cleaning diode history the intimate MS contact (HF etching or thermal pre-treatment at 700$^\circ$C) leads to the strongest Fermi level-pinning and to the highest spatial inhomogeneities without any relevant differences on SBH characteristics. It should be noted that during the thermal pre-treaments metallic impurities coming from the deposition chamber were introduced in the Ge substrate however in too low concentrations to affect the I-V forward characteristics. The insertion of a thin oxide interlayer (reference sample and pre-treatment at 400$^\circ$C) yielded to a depinning of the FL and to a more homogeneous contacts. While the nature of the oxide interlayer did not show significant differences on the SBH inhomogeneities it greatly affects the FL depinning. Ours results suggest that a pure $GeO_2$ oxide interlayer should present a better depinning efficiency than the native Ge oxide. Even for basic Schottky contacts on n-Ge, the electrical behaviors of the diodes are strongly modulated by the spatial inhomogeneities. Our work highlights the necessity of taking into account these inhomogenities in the interpretation of electrical characteristics. In particular the values deduced from the classic Richardson plots can be strongly inaccurate.\\

\ack

 We would like to thank M. Marteau (Pprime) for the thin film depositions and E. Simoen (imec) for the fruitful discussions.

\section*{References}

\bibliography{mapubli}

\providecommand{\newblock}{}
\begin{thebibliography}{10}
\expandafter\ifx\csname url\endcsname\relax
  \def\url#1{{\tt #1}}\fi
\expandafter\ifx\csname urlprefix\endcsname\relax\def\urlprefix{URL }\fi
\providecommand{\eprint}[2][]{\url{#2}}
% Bibliography created with iopart-num v2.1
% /biblio/bibtex/contrib/iopart-num

\bibitem{dimoulas07a}
Dimoulas A, Brunco D~P, Ferrari S, Seo J~W, Panayiotatos Y, Sotiropoulos A,
  Conard T, Caymax M, Spiga S, Fanciulli M, Dieker C, Evangelou E~K, Galata S,
  Houssa M and Heyns M~M 2007 {\em Thin Solid Films\/} {\bf 515} 6337--6343

\bibitem{kamata08MT}
Kamata Y 2008 {\em Mater. Today\/} {\bf 11} 30--38

\bibitem{brotzmann08}
Brotzmann S and Bracht H 2008 {\em J. Appl. Phys.\/} {\bf 103} 033508

\bibitem{kobayashi09}
Kobayashi M, Kinoshita A, Saraswat K, Wong H~S~P and Nishi Y 2009 {\em J. Appl.
  Phys.\/} {\bf 105} 023702

\bibitem{dubois04}
Dubois E and Larrieu G 2004 {\em J. Appl. Phys.\/} {\bf 96} 729

\bibitem{dimoulas09mrs}
Dimoulas A, Toriumi A and Mohney S~E 2009 {\em MRS Bull.\/} {\bf 34} 522

\bibitem{chi05}
Chi D~Z, Lee R~T~P, Chua S~J, Lee S~J, Ashok S and Kwong D~L 2005 {\em J. Appl.
  Phys.\/} {\bf 97} 113706

\bibitem{tsipas09}
Tsipas P and Dimoulas A 2009 {\em Appl. Phys. Lett.\/} {\bf 94} 012114

\bibitem{zhou08}
Zhou Y, Ogawa M, Han X and Wang K~L 2008 {\em Appl. Phys. Lett.\/} {\bf 93}
  202105

\bibitem{tung00}
Tung R~T 2000 {\em Phys. Rev. Lett.\/} {\bf 84} 6078--6081

\bibitem{zhou10}
Zhou Y, Han W, Wang Y, Xiu F, Zou J, Kawakami R~K and Wang K~L 2010 {\em Appl.
  Phys. Lett.\/} {\bf 96} 102103

\bibitem{nishimura07}
Nishimura T, Kita K and Toriumi A 2007 {\em Appl. Phys. Lett.\/} {\bf 91}
  123123

\bibitem{yamane10}
Yamane K, Hamaya K, Ando Y, Enomoto Y, Yamamoto K, Sadoh T and Miyao M 2010
  {\em Appl. Phys. Lett.\/} {\bf 96} 162104

\bibitem{werner91}
Werner J~H and G{\"u}ttler H~H 1991 {\em J. Appl. Phys.\/} {\bf 69} 1522

\bibitem{zhu00}
Zhu S, Detavernier C, Van~Meirhaeghe R~L, Cardon F, Ru G~P, Qu X~P and Li B~Z
  2000 {\em Solid-State Electron.\/} {\bf 44} 1807--1818

\bibitem{pakma2008}
Pakma O, Serin N, Serin T and Altindal S 2008 {\em J. Appl. Phys.\/} {\bf 104}
  014501

\bibitem{kavasoglu09}
Kavasoglu A~S, Tozlu C, Pakma O, Kavasoglu N, Ozden S, Metin B, Birgi O and
  Oktik S 2009 {\em J. Phys. D: Appl. Phys.\/} {\bf 42} 145111

\bibitem{husain09a}
Husain M~K, Li X~V and De~Groot C~H 2009 {\em Mater. Sci. Semicond. Process.\/}
  {\bf 11} 305

\bibitem{sun06}
Sun S, Sun Y, Liu Z, Lee D~I, Peterson S and Pianetta P 2006 {\em Appl. Phys.
  Lett.\/} {\bf 88} 021903

\bibitem{kamata08}
Kamata Y, Ino T, Koyama M and Nishiyama A 2008 {\em Appl. Phys. Lett.\/} {\bf
  92} 063512

\bibitem{prabhakaran00}
Prabhakaran K, Maeda F, Watanabe Y and Ogino T 2000 {\em Appl. Phys. Lett.\/}
  {\bf 76} 2244

\bibitem{dimoulas06}
Dimoulas A, Tsipas P, Sotiropoulos A and Evangelou E 2006 {\em Appl. Phys.
  Lett.\/} {\bf 89} 252110

\bibitem{lauwaert09}
Lauwaert J, Van~Gheluwe J, Vanhellemont J, Simoen E and Clauws P 2009 {\em J.
  Appl. Phys.\/} {\bf 105} 073707

\bibitem{auret06}
Auret F~D, Meyer W~E, Coelho S and Hayes M 2006 {\em Appl. Phys. Lett.\/} {\bf
  88} 242110--

\bibitem{rhoderick82}
Rhoderick E~H;~Williams R~H 1988 {\em Metal-semiconductor contacts\/}
  (Clarendon Press, Oxford)

\bibitem{kim09}
Kim H, Park C, Lee S and Kim D~W 2009 {\em J. Phys. D: Appl. Phys.\/} {\bf 42}
  055306

\bibitem{mtangi09}
Mtangi W, Auret F~D, Nyamhere C, Janse~van Rensburg P~J, Diale M and Chawanda A
  2009 {\em Physica B\/} {\bf 404} 1092--1096

\bibitem{biber03}
Biber M 2003 {\em Physica B\/} {\bf 325} 138--148

\bibitem{wenckstern06}
von Wenckstern H, Biehne G, Rahman R, Hochmuth H, Lorenz M and Grundmann M 2006
  {\em Appl. Phys. Lett.\/} {\bf 88} 092102

\bibitem{pelissier07}
Pelissier B, Kambara H, Godot E, Veran E, Loup V, Bensahel D and Joubert O 2007
  {\em Semiconductor International\/} {\bf 30} 77--83

\bibitem{jerl}
Gerl M and Issi J 1997 {\em Physique des mat\'eriaux, tome 8\/} (Presses
  polytechniques et universitaire romandes)

\bibitem{zou07}
Zou X, Xu J~P, Li C~X and Lai P~T 2007 {\em Appl. Phys. Lett.\/} {\bf 90}
  163502

\bibitem{simoen08d}
Simoen E, Opsomer K, Claeys C, Maex K, Detavernier C, Van~Meirhaeghe R~L and
  Clauws P 2008 {\em Solid State Phenom.\/} {\bf 131-133} 47--52

\bibitem{simoen08}
Simoen E, Opsomer K, Claeys C, Maex K, Detavernier C, Van~Meirhaeghe R~L and
  Clauws P 2008 {\em J. Appl. Phys.\/} {\bf 104} 023705

\bibitem{markevich04}
Markevich V, Peaker A, Litvinov V, Emtsev V and Murin L 2004 {\em J. Appl.
  Phys.\/} {\bf 95} 4078

\bibitem{pearton82}
Pearton S~J 1982 {\em Solid State Electron.\/} {\bf 25} 305--311

\bibitem{simoen06}
Simoen E, Opsomer K, Claeys C, Maex K, Detavernier C, Van~Meirhaeghe R~L,
  Forment S and Clauws P 2006 {\em Appl. Phys. Lett.\/} {\bf 88} 183506

\bibitem{auret08}
Auret F~D, Coelho S~M~M, Hayes M, Meyer W~E and Nel J~M 2008 {\em Phys. Status
  Solidi (A)\/} {\bf 205} 159--161

\bibitem{fage00}
Fage-Pedersen J, Nylandsted~Larsen A and Mesli A 2000 {\em Phys. Rev. B\/} {\bf
  62} 10116--10125

\bibitem{claeys07}
Claeys C and Simoen E 2007 {\em Germanium-based technologies: from materials to
  devices\/} (Elsevier Science)

\bibitem{simoen07}
Simoen E, Claeys C, Sioncke S, Steenbergen J, Meuris M, Forment S, Vanhellemont
  J, Clauws P and Theuwis A 2007 {\em J. Mater. Sci.: Mater. Electron.\/} {\bf
  18} 799--804

\bibitem{werner88}
Werner J~H 1988 {\em Appl. Phys. A: Mater. Sci. Process.\/} {\bf 47} 291--300

\bibitem{sze07}
Sze S~M and Ng K~K 2007 {\em Physics of semiconductor devices\/} (Wiley
  Interscience)

\bibitem{yao06}
Yao H~B, Chi D~Z, Li R, Lee S~J and Kwong D~L 2006 {\em Appl. Phys. Lett.\/}
  {\bf 89} 242117

\bibitem{werner93}
Werner J~H and G{\"u}ttler H~H 1993 {\em J. Appl. Phys.\/} {\bf 73} 1315

\bibitem{auvergne74}
Auvergne D, Camassel J, Mathieu H and Cardona M 1974 {\em Phys. Rev. B\/} {\bf
  9} 5168--5177

\bibitem{lautenschlager85}
Lautenschlager P, Allen P~B and Cardona M 1985 {\em Phys. Rev. B\/} {\bf 31}
  2163--2171

\bibitem{rivillon05}
Rivillon S, Chabal Y, Amy F and Kahn A 2005 {\em Appl. Phys. Lett.\/} {\bf 87}
  253101

\bibitem{tung92}
Tung R~T 1992 {\em Phys. Rev. B\/} {\bf 45} 13509--13523

\bibitem{li09}
Li X~V, Husain M~K, Kiziroglou M and de~Groot C~H 2009 {\em Microelectron.
  Eng.\/} {\bf 86} 1599--1602

\end{thebibliography}

\end{document}